\documentclass[webversion]{usetex-v1}

\usepackage{cite}	
\usepackage{url}
\usepackage{xspace}
\usepackage{graphicx}

\long\def\com#1{}

\long\def\xxx#1{{\bf XXX: }{\small [#1]}}

\begin{document}

\title{\vspace{-1em}
	Icebergs in the Clouds: 
	the {\em Other} Risks of Cloud Computing}

\author{
	\authname{Bryan Ford}
	\authaddr{Yale University}
	\authurl{\url{http://bford.info/}}
	\vspace{-2em}
}

\maketitle

\begin{abstract}

Cloud computing is appealing from management and efficiency perspectives,
but brings risks both known and unknown.
Well-known and hotly-debated information security risks,
due to software vulnerabilities, insider attacks, and side-channels for example,
may be only the ``tip of the iceberg.''
As diverse, independently developed cloud services
share ever more fluidly and aggressively multiplexed hardware resource pools,
unpredictable interactions between load-balancing and other reactive mechanisms
could lead to dynamic instabilities or ``meltdowns.''
Non-transparent layering structures,
where alternative cloud services may appear independent
but share deep, hidden resource dependencies,
may create unexpected and potentially catastrophic failure correlations,
reminiscent of financial industry crashes.
Finally,
cloud computing exacerbates already-difficult
digital preservation challenges,
because only the provider of a cloud-based application or service
can archive a ``live,'' functional copy
of a cloud artifact and its data
for long-term cultural preservation.
This paper explores these largely unrecognized risks,
making the case that we should study them {\em before}
our socioeconomic fabric becomes inextricably dependent on
a convenient but potentially unstable computing model.

\end{abstract}

\section{Introduction}

\com{ work in refs: ~\cite{raghavendra08power,mogul06emergent}.
Also, John Edwards who contacted me by E-mail about police \& fire depts.
Also, guy from IBM at the NSF workshop who mentioned encountering
instability issues in IBM's internal cloud... 
Can't remember who it was though;
maybe it'll come out in the workshop notes?
Also: Schapira's paper:
http://www.cs.yale.edu/homes/schapira/jsw-ics11.pdf (full),
http://www.cs.yale.edu/homes/schapira/podc171b-jaggard.pdf (summary).

Related to future ``multi-cloud/multi-provider'' trend:
Xen Blanket, EuroSys '12~\cite{williams12xen}
}

Attractive features and industry momentum
make cloud computing appear destined
to be the next dominant computing paradigm.
Cloud computing is appealing
due to the convenience of central management
and the elasticity of resource provisioning.
Moving critical information infrastructure to the cloud
also presents risks, however, some of which are well-known
and already hot research topics.
The much-discussed challenge of ensuring the privacy
of information hosted in the cloud, for example~\cite{
	gellman09privacy},
has resulted in an emerging breed
of ``cloud-hardened'' virtualization hardware~\cite{
	keller10nohype}
and security kernels~\cite{zhang11cloudvisor}.
Similarly, the challenge of ensuring high availability in the cloud
has in part fueled recent research on
robust data center networking~\cite{
	wang10r3,raiciu11improving}.

This paper assumes that
a large fraction of the computing industry is,
for better or worse,
``moving to the cloud,''
and that
current research addressing
the immediate information security risks is well underway
and will (eventually) succeed.
Setting aside these known challenges, therefore,
this paper attempts to identify and focus on several
{\em less} well-understood---and perhaps less ``imminent''---risks
that {\em may} emerge from the shift to cloud computing.
In particular,
this paper addresses:
(1) 
	{\bf stability} risks due to unpredictable interactions between
	independently developed but interacting cloud computations;
(2)	{\bf availability} risks due to non-transparent layering
	resulting in hidden failure correlations; and
(3)	{\bf preservation} risks due to the unavailability of
	a cloud service's essential code and data outside of the provider.

This paper is speculative and forward-looking;
the author cannot yet offer definitive evidence that any of these risks
{\em will} fully materialize or become vitally important,
but rather can offer only informal arguments and anecdotal evidence
that these risks {\em might} become important issues.
The above list is also probably incomplete:
it is likely that other important risks will emerge only
as the industry continues its shift to the cloud.
Nevertheless,
I argue that it is worth proactively investigating longer-term risks
such as these before they are certain or imminent,
as the stakes may be high.
Further, once any of these risks {\em do} become important,
it may be too late to reconsider or slow
the movement of critical infrastructure to the cloud,
or to rethink the architecture of important cloud infrastructure or services
once they are already perceived as ``mature'' in the industry.

Section~\ref{sec-stab} addresses stability risks,
Section~\ref{sec-avail} explores availability risks, and
Section~\ref{sec-pres} explores preservation risks.
Section~\ref{sec-sol} briefly points out a few possible research directions
in which solutions might be found---%
though this paper cannot and does not pretend to offer ``answers.''
Finally,
Section~\ref{sec-concl} concludes.

\section{Stability Risks from Interacting Services}
\label{sec-stab}

Cloud services and applications increasingly build atop one another
in ever more complex ways,
such as cloud-based advertising or mapping services
used as components in other, higher-level cloud-based applications,
all of these building on computation and storage infrastructure
offered by still other providers.
Each of these interacting, codependent services and infrastructure components
is often implemented, deployed, and maintained independently
by a single company that, for reasons of competition,
shares as few details as possible about the internal operation of its services.
The resource provisioning and moment-by-moment operation of each service
is often managed by dynamic, reactive control processes
that constantly monitor the behavior of customer load,
internal infrastructure,
and other component services,
and implement complex proprietary policies
to optimize the provider's cost-benefit ratio.

Each cloud service's control loop
may change the service's externally visible behavior,
in policy-specific ways,
based on its neighboring services' behavior,
creating cyclic control dependencies
between interacting cloud services.
These dependency cycles may lead to unexpected feedback and instability,
in much the way that policy-based routing in BGP
is already known to lead to instability or ``route flapping''
in the much more restricted ``control domain''
of Internet routing~\cite{varadhan00persistent,griffin02stable}.

\begin{figure}[t]
\centering
\includegraphics[width=0.45\textwidth]{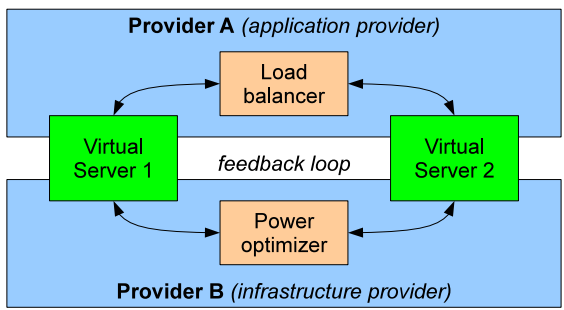}
\caption{Example instability risk from unintended coupling
	of independently developed reactive controllers}
\label{fig-stab}
\end{figure}

To illustrate this risk,
we consider a minimalistic, perhaps contrived,
but hopefully suggestive example
in Figure~\ref{fig-stab}.
Application provider $A$ develops and deploys a cloud-based application,
which runs on virtual compute and storage nodes
from infrastructure provider $B$.
For simplicity,
assume $A$ leases two virtual nodes from $B$,
and dynamically load-balances incoming requests
across the web/application servers running on these nodes.
Assume $A$'s load balancer operates in a control loop with a 1-minute period:
after each minute it evaluates each server's current load
based on that server's response time statistics during the past minute,
and shifts more traffic during the next minute to the less-loaded server.
Assume that $A$'s load shifting algorithm is well-designed and stable
assuming the servers in the pool behave consistently over time,
like dedicated physical servers would.

Unbeknownst to $A$, however, suppose
$B$ also runs a control loop,
which attempts to optimize the power consumption of its physical servers
by dynamically adjusting the servers' clock rates based on load.
This control loop also happens to have a 1-minute period:
after each minute, $B$'s controller measures each CPU core's utilization
during the past minute,
then reduces the core's voltage and speed if the core was underutilized
or increases voltage and speed if the core was overutilized.
Again, assume that $B$'s controller is well-designed
and stable assuming that the servers' load stays relatively constant
or varies independently of $B$'s control actions.

Although both $A$'s and $B$'s control loops would be stable if operating alone,
by the misfortune of their engineers (independently)
picking similar control loop periods,
the combination of the two control loops may risk a positive feedback loop.
Suppose during one minute the load is slightly imbalanced
toward virtual server 1,
and the two control loops' periods happen to be closely aligned;
this will happen sooner or later in the likely event their clocks
run at slightly different rates.
$A$'s load balancer notices this and shifts some load away from the node
in the next minute,
while $B$'s power optimizer notices the same thing
and increases the node's voltage and clock speed.
While either of these actions alone would lead toward convergence,
the two in combination cause overcompensation:
during the next minute, server 1 becomes {\em more} underutilized
than it was overutilized in the previous minute.
The two controllers each compensate with a stronger action---%
a larger shift of traffic back to server 1 by $A$
and a larger decrease in voltage and clock speed by $B$---%
causing a larger swing the next minute.
Soon all incoming load is oscillating between the two servers,
cutting the system's overall capacity in half---%
or worse, if more than two servers are involved.

This simplistic example
might be unlikely to occur
in exactly this form on real systems---%
or might be quickly detected and ``fixed''
during development and testing---%
but it suggests a general risk.
When multiple cloud services independently
attempt to optimize their own operation
using control loops that both monitor, and affect,
the behavior of upstream, downstream, or neighboring cloud services,
it is hard to predict the outcome:
we might well risk deploying a combination of control loops
that behaves well ``almost all of the time,''
until the emergence of the rare, but fatal, cloud computing equivalent
of the Tacoma Narrows Bridge~\cite{billah91resonance,mckenna99large}.

Comparable forms of ``emergent misbehavior''
have been observed in real computing systems
outside of the cloud context~\cite{mogul06emergent},
and some work has studied the challenge of coordinating and stabilizing
multiple interacting control loops,
such as in power management~\cite{raghavendra08power}.
Current approaches to solving or heading off such
instability risks, however,
generally assume that {\em some} single engineer or company
has complete information about, and control over,
all the interacting layers and their control loops.
The cloud business model undermines this design assumption,
by incentivizing providers {\em not} to share with each other
the details of their resource allocation and optimization algorithms---%
crucial parts of their ``secret sauce''---%
that would be necessary to analyze or ensure the stability
of the larger, composite system.

\com{
While it is unclear to what extent this is or will be a problem
or whether a general and realistic solution exists,
Section~\ref{sec-stab} explores one potential approach,
based on a variation of the same labeling technique
with which we hope to address the timing channel problem above.
}

\section{Risks of Hidden Failure Correlations}
\label{sec-avail}

Ensuring high availability is usually
a high priority for cloud infrastructure and services,
and state replication and fault tolerance mechanisms
is the focus of much industry and research attention.
Most of this attention is focused {\em within} a particular cloud service,
however.
In addition to the stability risks discussed above,
interactions between multiple interdependent cloud services
could lead to availability risks not yet addressed in mainstream research,
where hardware infrastructure interdependencies
hidden by proprietary business relationships
can lead to unexpected failure correlations.

\begin{figure}[t]
\centering
\includegraphics[width=0.45\textwidth]{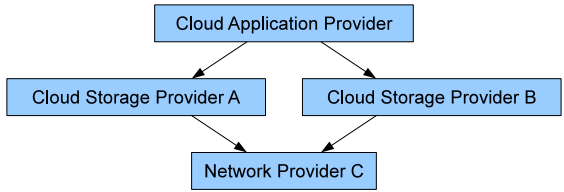}
\caption{Cloud service stack illustrating risks of correlated failures
	due to hidden service interdependencies}
\label{fig-stack}
\end{figure}

As another contrived but illustrative example,
consider the ``cloud service stack'' in Figure~\ref{fig-stack}.
The provider at the top offers a cloud-based application
intended to offer mission-critical reliability.
To ensure this reliability, the application
replicates all critical application state
across the storage services provided by
two nominally-independent cloud storage providers, A and B,
each of which in turn provides storage at multiple geographic sites
with separate network connectivity at each site.

Unbeknownst to the application provider, however,
each storage provider obtains its network connections
from a common underlying network provider, C.
The application's access to its critical storage
proves highly reliable
as long as provider C operates normally.
If provider C encounters a rare disaster or administrative glitch, however---%
or enters a dispute
with another top-tier network provider~\cite{bray05dispute}---%
the mission-critical application may suddenly lose connectivity
to {\em both} of its critical storage repositories.
This correlated failure results from the shared dependencies on C
being hidden by the proprietary business relationships
through which the application provider obtains services from A and B.

As the cloud computing industry matures
and produces ever more complex cloud-based services,
it seems inevitable that the depth and complexity of inter-service relationships
will continue to explode,
which may create unpredictable availability risks
due to ever more subtle cross-layer interdependencies,
of which the above example is merely the most simplistic representative.
Furthermore, one of the fundamental attractions of cloud computing
is that it makes computing infrastructure, services, and applications
into generic, almost arbitrarily ``fungible'' resources
that can be bought, sold, and resold
as demanded by business objectives~\cite{williams12xen}.

It does not seem far-fetched to predict that cloud services will arise
that represent a thin veneer over, or ``repackaging'' of,
other services or combinations of services:
e.g., businesses that resell, trade, or speculate on
complex cocktails or ``derivatives'' of more basic cloud resources and services,
much like the modern financial and energy trading industries operate.
If this prediction bears out,
the cloud services industry could similarly start yielding
speculative bubbles and occasional large-scale failures,
due to ``overly leveraged'' composite cloud services
whose complex interdependencies hide correlated failure modes
that do not become apparent until the bubble bursts catastrophically---%
perhaps not wholly unlike the causes of the recent financial meltdown
or the earlier Enron energy bubble~\cite{healy03fall}.
Once again, while this risk is pure speculation at this point,
it seems worth taking seriously and exploring in advance.
\com{
Section~\ref{sec-avail} briefly outlines one possible technical approach
for heading off such risks.
}


\section{Digital Preservation Risks}
\label{sec-pres}

The final risk considered here is more long-term.
With the tremendous economic momentum toward
cloud-based and cloud-dependent applications and services,
it appears inevitable that these cloud-based ``digital artifacts''
will soon represent a considerable and ever-increasing component
of our social and cultural heritage.
In 100 years, however,
will today's culturally important
cloud-based digital artifacts
still be available in a historically accurate form---%
or in any form?

A physical book has an inherent {\em decentralized archivability} property.
In order to make money on a book,
its author or publisher must make complete copies available to customers.
Customers in turn are free to---and cannot effectively be prevented from---%
independently storing books for any amount of time,
relocating copies to a safe long-term repository (e.g., a library),
copying them to other media as the original media deteriorates, etc.

Preservation of digital works
presents many known challenges---%
principally the faster deterioration or obsolescence of electronic media,
and the obsolescence of computing environments
needed to interpret old data formats~\cite{
	rothenberg99avoiding,bearman99reality,maniatis05lockss}.
Yet despite these known challenges,
traditional software and associated documents
stored on a floppy or hard disk, USB stick,
or even a ``cloud drive'' holding raw files,
still has the same {\em decentralized archivability} property of a book.
The vendor of a traditional software application or digital document
must, in order to make money,
make essentially {\em complete} copies available to customers,
and these customers can work in an arbitrarily decentralized fashion
using their own resources
to preserve digital works deemed worth saving.

Cloud-based applications and services, however,
completely eliminate this property of decentralized archivability.
Unlike users of Microsoft Office,
users of Google Search or Maps never gain access to
anything remotely resembling a ``complete copy'' of
the entire digital artifact represented by the Google Search or Maps service.
At most, users might save the results
of particular queries or interactions.
Unlike players of Doom,
players of World of Warcraft (WoW)
cannot independently archive and preserve a copy of the WoW universe---%
or even a small portion of interest---%
because the provider of the cloud-based application need not,
and typically does not,
make publicly available 
the server-side software and data comprising the service.

Given the number of scholarly papers written
on the technological and social implications of each,
it would be hard to argue that Google Search and WoW
do not represent a historically significant digital artifacts.
Yet given the rate that Google and Blizzard evolve their services
to compete more effectively in the search and gaming markets, respectively,
it is almost certain that ten years from now,
no one outside these companies---%
perhaps not even anyone {\em inside} them---%
will be able to reproduce a faithful, functioning copy
of the Google Search or WoW service {\em as it exists today}.
In 100 years,
these services will probably have evolved beyond recognition,
assuming they survive at all.

If today's digital archivists do their jobs well,
in 100 years we will be able to run today's Microsoft Word or play Doom
(in an emulator if necessary)---%
but nothing today's digital archivists can do
will preserve historically relevant snapshots of today's cloud-based services,
because the archivists never even get access
to a ``complete'' snapshot for preservation.

The historical record of today's Google Search or WoW
will consist merely of second-hand accounts:
articles written about them, saved search queries or screen shots,
captured videos of particular WoW games, etc.
While better than nothing,
such second-hand accounts would not suffice
for future historians to answer questions such as:
``How did the focus or breadth of search results for interesting queries
evolve over the last 10 or 100 years?''
Or,
``How did social-interaction and player-reward mechanisms
change in MMOGs historically?''

These particular examples may or may not seem interesting or important,
but the point is that {\em we don't know}
what future historians or social scientists will deem important
about today's world.
As more of today's culture shifts to the cloud,
our failure to preserve our cloud-based digital artifacts
could produce a ``digital dark age'' far more opaque and impenetrable
to future generations
than what media or OS obsolescence alone will produce.

\com{
While this problem is clearly not likely to present a simple or easy solution,
Section~\ref{sec-preserv} proposes one way
in which cloud infrastructure could be evolved
so as to give all stakeholders in a cloud service---%
not just the provider but also customers and archivists---%
some ability to preserve functioning snapshots of cloud services.
}

\com{
- incentives don't encourage developer to make archival copies available
or preserve them; economic incentives normally only make the ``latest version''
interesting, and encourage constant upgrades to keep up with the competition.

- even if app developer is willing to make the code and data available
for archival storage and historical use,
the relevant datasets may be huge,
widely distributed across thousands of machines in multiple data centers,
and constantly-changing:
who will pay for the cost of storing snapshots of these huge artifacts?
Customers and librarians probably can't afford to, at least individually.

- app developer may be willing to make only subsets of their dataset
available to a particular stakeholder:
e.g., only the portion of map data requested by a customer at a particular time.
}

\section{In Search of Possible Solutions}
\label{sec-sol}

This paper cannot hope to---and makes no attempt to---%
offer solutions or answers to the problems outlined above.
Instead, we merely conjecture at a few potential directions
in which solutions {\em might} be found.

\paragraph{Stabilizing Cloud Services:}
One place we might begin to study
stability issues between interacting cloud services, and potential solutions,
is the extensive body of work on the unexpected inter-AS
(Autonomous System) interactions frequently observed in BGP routing~\cite{
	varadhan00persistent,griffin02stable}.
In particular,
the ``dependency wheel'' model, useful for reasoning about BGP policy loops,
seems likely to generalize to higher-level control loops in the cloud,
such as load balancing policies.
Most of the potential {\em solutions} explored so far in the BGP space,
however, appear largely specific to BGP---or at least to routing---%
and may have to be rethought ``fram scratch''
in the context of more general, higher-level cloud services.

Beyond BGP,
classic control theory may offer a broader source of inspiration
for methods of understanding and ensuring cloud stability.
Most conventional control-theoretic techniques, however,
are unfortunately constructed from the assumption
that some ``master system architect'' can control or at least describe
all the potentially-interacting control loops in a system to be engineered.
The cloud computing model violates this assumption at the outset
by juxtaposing many interdependent, reactive control mechanisms
that are by nature {\em independently} developed,
and are often the proprietary and closely-guarded
business secrets of each provider.

\paragraph{Deep Resource (In)Dependence Analysis:}

\begin{figure*}[t]
\centering
\includegraphics[width=0.74\textwidth]{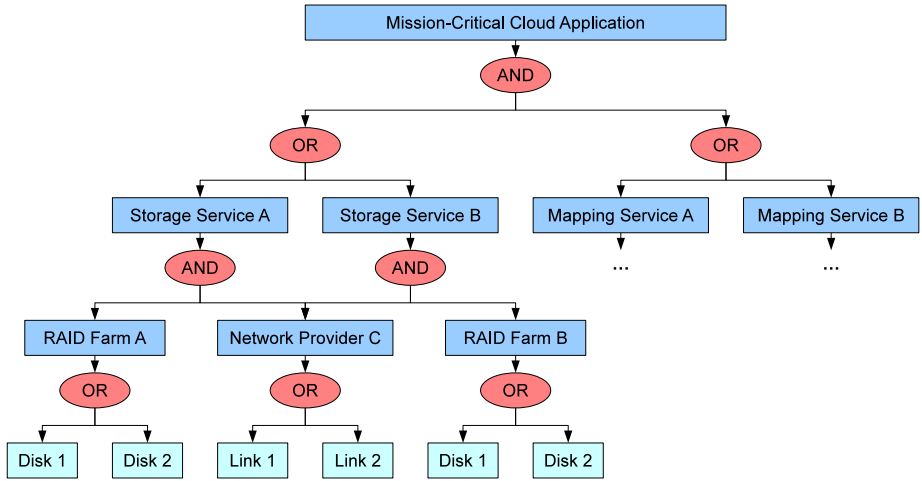}
\caption{AND/OR Graph Representing Service Composition
		and Infrastructure Dependencies}
\label{fig-andor}
\end{figure*}

The availability risks discussed in Section~\ref{sec-avail}
result from the fact that cloud service and infrastructure providers
usually do not reveal the deep dependency structure underlying their services.
The key to this risk is the non-transparency of the dependency graph:
the application provider in Figure~\ref{fig-stack}
{\em does not know} that both A and B depend on the same network provider C,
resulting in hidden failure correlations.
Supposing the providers were to make these dependencies
visible in an explicit dependency graph, however,
we might be able to estimate {\em actual} dependence or independence
between different services or resources for reliability analysis.

Hardware design techniques
such as fault tree analysis~\cite{vesely81fault,bedford01probabilistic}
may offer some tools that could be adapted
to the purpose of reasoning about
cloud service and infrastructure dependencies.
Consider for example a simplistic
AND/OR resource dependency graph, shown in Figure~\ref{fig-andor}.
AND nodes reflect design composition and hence conjunctive dependency:
{\em all} components underneath an AND node must function correctly
in order for the component above to operate.
OR nodes reflect design redundancy and hence disjunctive dependency:
if {\em any} component underneath the OR node operates,
the dependent component above the OR will operate.
Given such a graph,
annotated with expected failure rates,
one might compute or estimate a system's {\em effective} reliability
after accounting for unanticipated common dependencies,
such as Network Provider C in the example.

Cloud providers may be reluctant to release
detailed dependency information publicly for business reasons,
but might willing release it to a trusted third party,
such as an organization analogous to Underwriters Laboratories (UL)
offering cloud reliability analysis services.
More ambitiously,
cloud providers might leverage TPM-attested, IFC-enforcing kernels~\cite{
	zeldovich06making}
to exchange and analyze dependency graph information,
without allowing proprietary information to ``leak'' beyond this analysis.

\paragraph{Preserving Cloud Artifacts:}
Enabling the long-term preservation of cloud artifacts will require solving
both incentive problems and technical challenges.

In a cloud-based computing model,
application and service providers currently need not,
and have little incentive to,
make publicly available all the software and data underlying the service
that would be necessary for accurate historical preservation.
Competition encourages providers to closely guard
the ``secret sauce'' underlying their products.
This incentive has long led traditional software vendors
to release their software only in binary form---%
often with deliberate obfuscation to thwart analysis---%
but only the cloud model frees the vendor entirely
from the need to release their code in {\em any} form
directly executable by the customer.
Solving this incentive problem will likely require
social, commercial, and/or governmental incentives
for providers to make their cloud-based artifacts preservable in some way.

On the technical side,
cloud-based services often rely on enormous, frequently-changing datasets,
such as the massive distributed databases underlying Google Search or Maps
or an MMOG's virtual world.
Even if willing,
it might be impractically costly for providers
to ship regular snapshots of their entire datasets to digital archivists---%
even well-provisioned ones such as the Library of Congress---%
not to mention costly for receiving archivists
to do anything with such enormous snapshots beyond saving the raw bits.
A more practical approach may be for providers themselves to be responsible
for saving historical snapshots in the short term,
using standard copy-on-write cloning
and deduplicated storage technologies for efficiency~\cite{
        santry99elephant,quinlan02venti}.
After some time period, say 5--10 years,
a select subset of these historical snapshots might then be transferred
to external archives for long-term preservation,
at considerably reduced cost-per-bit
in terms of both network bandwidth and storage
due to intervening technological evolution.

Any solution would need to address many other challenges,
such as
ensuring the durability and integrity of online digital archives~\cite{
	maniatis05lockss}
and the honesty of their providers~\cite{shah11auditing},
maintaining information security of sensitive data
in snapshots of cloud-based artifacts,
and preserving artifacts' practical usability in addition to their raw bits,
but we leave these issues to future work.

\section{Conclusion}
\label{sec-concl}

While the cloud computing model is promising and attractive in many ways,
the author hopes that this paper has made the case
that the model may bring risks beyond obvious information security concerns.
At the very least,
it would be prudent for us to study some of these risks
{\em before} our socioeconomic system becomes completely and irreversibly
dependent on a computing model whose foundations
may still be incompletely understood.

\subsection*{Acknowledgments}

Jeff Mogul, Michael Schapira, John Edwards,
and the anonymous HotCloud reviewers offered valuable feedback
on early drafts of this paper.
This research was sponsored by the NSF
under grant CNS-1149936.

\bibliography{os,net,soc}

\begin{thebibliography}{10}

\bibitem{bearman99reality}
David Bearman.
\newblock Reality and chimeras in the preservation of electronic records.
\newblock {\em D-Lib Magazine}, 5(4), April 1999.

\bibitem{bedford01probabilistic}
Tim Bedford and Roger Cooke.
\newblock {\em Probabilistic Risk Analysis: Foundations and Methods}.
\newblock Cambridge University Press, 2001.

\bibitem{billah91resonance}
K.~Yusuf Billah and Robert~H. Scanlan.
\newblock Resonance, {Tacoma Narrows} bridge failure, and undergraduate physics
  textbooks.
\newblock {\em American Journal of Physics}, 59(2), February 1991.

\bibitem{bray05dispute}
Hiawatha Bray.
\newblock Dispute threatens to snarl {Internet}.
\newblock {\em Boston Globe}, October 2005.

\bibitem{gellman09privacy}
Robert Gellman.
\newblock Privacy in the clouds: Risks to privacy and confidentiality from
  cloud computing.
\newblock {\em World Privacy Forum}, pages 1--16, 2009.

\bibitem{griffin02stable}
Timothy~G. Griffin, F.~Bruce Shepherd, and Gordon Wilfong.
\newblock The stable paths problem and interdomain routing.
\newblock {\em Transactions on Networking}, 10(2), April 2002.

\bibitem{healy03fall}
Paul~M. Healy and Krishna~G. Palepu.
\newblock The fall of {Enron}.
\newblock {\em Journal of Economic Perspectives}, 17(2):3--26, 2003.

\bibitem{keller10nohype}
Eric Keller et~al.
\newblock {NoHype}: Virtualized cloud infrastructure without the
  virtualization.
\newblock In {\em \bibconf[37th]{ISCA}{International Symposium on Computer
  Architecture}}, June 2010.

\bibitem{maniatis05lockss}
Petros Maniatis et~al.
\newblock The {LOCKSS} peer-to-peer digital preservation system.
\newblock {\em \bibbrev{TOCS}{Transactions on Computer Systems}}, 23(1):2--50,
  2005.

\bibitem{mckenna99large}
P.J. McKenna.
\newblock Large torsional oscillations in suspension bridges revisited: Fixing
  an old approximation.
\newblock {\em The American Mathematical Monthly}, 106(1), January 1999.

\bibitem{mogul06emergent}
Jeffrey~C. Mogul.
\newblock Emergent (mis)behavior vs. complex software systems.
\newblock In {\em \bibconf[1st]{EuroSys}{ACM SIGOPS/EuroSys European Conference
  on Computer Systems}}, 2006.

\bibitem{quinlan02venti}
S.~Quinlan and S.~Dorward.
\newblock {Venti}: a new approach to archival storage.
\newblock In {\em \bibconf[1st]{FAST}{USENIX Conference on File and Storage
  Technologies}}, 2002.

\bibitem{raghavendra08power}
Ramya Raghavendra, Parthasarathy Ranganathan, Vanish Talwar, Zhikui Wang, and
  Xiaoyun Zhu.
\newblock No ``power'' struggles: coordinated multi-level power management for
  the data center.
\newblock In {\em \bibconf[13th]{ASPLOS}{International Conference on
  Architectural Support for Programming Languages and Operating Systems}},
  pages 48--59, 2008.

\bibitem{raiciu11improving}
Costin Raiciu et~al.
\newblock Improving datacenter performance and robustness with {Multipath TCP}.
\newblock In {\em \bibbrev{SIGCOMM}{ACM SIGCOMM}}, August 2011.

\bibitem{rothenberg99avoiding}
Jeff Rothenberg.
\newblock {\em Avoiding Technological Quicksand: Finding a Viable Technical
  Foundation for Digital Preservation}.
\newblock Council on Library and Information Resources, January 1999.

\bibitem{santry99elephant}
Douglas~S. Santry et~al.
\newblock Deciding when to forget in the {Elephant} file system.
\newblock In {\em \bibconf[17th]{SOSP}{ACM Symposium on Operating Systems
  Principles}}, December 1999.

\bibitem{shah11auditing}
Mehul~A. Shah, Mary Baker, Jeffrey~C. Mogul, and Ram Swaminathan.
\newblock Auditing to keep online storage services honest.
\newblock In {\em \bibconf[11th]{HotOS}{Workshop on Hot Topics in Operating
  Systems}}, May 2007.

\bibitem{varadhan00persistent}
Kannan Varadhan, Ramesh Govindan, and Deborah Estrin.
\newblock Persistent route oscillations in inter-domain routing.
\newblock {\em Computer Networks}, 32(1):1--16, January 2000.

\bibitem{vesely81fault}
W.E. Vesely, F.F. Goldberg, N.H. Roberts, and D.F. Haasl.
\newblock {\em Fault Tree Handbook}.
\newblock U.S. Nuclear Regulatory Commission, January 1981.

\bibitem{wang10r3}
Ye~Wang et~al.
\newblock R3: Resilient routing reconfiguration.
\newblock In {\em \bibbrev{SIGCOMM}{ACM SIGCOMM}}, August 2010.

\bibitem{williams12xen}
Dan Williams, Hani Jamjoom, and Hakim Weatherspoon.
\newblock The {Xen-Blanket}: Virtualize once, run everywhere.
\newblock In {\em \bibconf{EuroSys}{European Conference on Computer Systems}},
  April 2012.

\bibitem{zeldovich06making}
Nickolai Zeldovich et~al.
\newblock Making information flow explicit in {HiStar}.
\newblock In {\em \bibconf[7th]{OSDI}{USENIX Symposium on Operating Systems
  Design and Implementation}}, November 2006.

\bibitem{zhang11cloudvisor}
Fengzhe Zhang, Jin Chen, Haibo Chen, and Binyu Zang.
\newblock {CloudVisor}: Retrofitting protection of virtual machines in
  multi-tenant cloud with nested virtualization.
\newblock In {\em \bibconf[23rd]{SOSP}{{ACM} Symposium on Operating Systems
  Principles (SOSP)}}, October 2011.

\end{thebibliography}
\bibliographystyle{plain}

\end{document}